\documentclass[twocolumn,english,aps,prx,notitlepage,superscriptaddress]{revtex4-2}
\usepackage[T1]{fontenc}
\usepackage[latin9]{inputenc}
\usepackage{graphicx}
\usepackage{amsmath}
\usepackage{float}
\usepackage[colorlinks = true,
linkcolor = blue,
urlcolor  = blue,
citecolor = blue,
anchorcolor = blue]{hyperref}
\usepackage[usenames, dvipsnames]{color}

\usepackage{tikz}
\usetikzlibrary{shapes.geometric, arrows}

\makeatletter
\usepackage{babel}
\usepackage{float}
\usepackage{appendix}

\usepackage{tabularx}
\newcolumntype{C}{>{\centering\arraybackslash}X}

\makeatother

\begin{document}

\title{Tutorial: Extracting entanglement signatures from neutron spectroscopy}

\author{Allen Scheie}
\email{scheie@lanl.gov}
\affiliation{MPA-Q, Los Alamos National Laboratory, Los Alamos, New Mexico 87545, USA}

\author{Pontus Laurell}
\affiliation{Department of Physics and Astronomy, University of Missouri,
	Columbia, Missouri 65211, USA}

\author{Wolfgang Simeth}
\affiliation{MPA-Q, Los Alamos National Laboratory, Los Alamos, New Mexico 87545, USA}

\author{Elbio Dagotto}
\affiliation{Department of Physics and Astronomy, University of Tennessee, Knoxville, Tennessee 37996, USA}
\affiliation{Materials Science and Technology Division, Oak Ridge National Laboratory, Oak Ridge, Tennessee 37831, USA}

\author{D. Alan Tennant}
\affiliation{Department of Physics and Astronomy, University of Tennessee, Knoxville, Tennessee 37996, USA}
\affiliation{Department of Materials Science and Engineering, University of Tennessee, Knoxville, Tennessee 37996, USA}

\date{\today}

\begin{abstract}

This tutorial is a pedagogical introduction to recent methods of computing quantum spin entanglement witnesses from spectroscopy, with a special focus on neutron scattering on quantum spin systems. We offer a brief introduction to the concepts and equations, define a data analysis protocol, and discuss the interpretation of three entanglement witnesses: one-tangle, two-tangle, and Quantum Fisher Information. We also discuss practical experimental considerations, and give three examples of extracting entanglement witnesses from experimental data: Copper Nitrate, KCuF$_3$, and NiPS$_3$.  

\end{abstract}

\maketitle


\section{Introduction}

In solid state materials, quantum entanglement between electrons can drive systems into exotic states, including superconducting, quantum critical, quantum spin liquid, and fractional quantum states \cite{keimer2017physics}.  
Historically, the presence of highly entangled quantum states in condensed matter has been inferred by comparing materials to theoretical models. However, recent work has shown that spectroscopy can be used to directly measure quantum entanglement in materials in a model-independent fashion \cite{PhysRevResearch.2.043329,PhysRevB.103.224434,PhysRevLett.127.037201,Hauke2016,hales2023witnessing,Scheie2024_KYS,fang2024amplified,ren2024witnessing}. This allows detection of a highly entangled quantum state without recourse to detailed theoretical modeling (which is often prohibitively difficult). 

A detailed review of these methods has been recently given in Ref. \cite{PontusEntanglementReview}. This tutorial is a pedagogical guide to extracting quantum entanglement from spectroscopy with a special focus on neutron scattering. 
The outline for this tutorial is as follows: section \ref{sec:Definitions} defines key terms and equations, 
section \ref{sec:Protocol} outlines a data analysis protocol for extracting entanglement witnesses,
section \ref{sec:Interpretation} discusses the interpretation of spectroscopic quantum entanglement witnesses, 
section \ref{sec:ExperimentPlanning} discusses considerations for planning a neutron scattering experiment to extract entanglement, and finally section \ref{sec:Examples} gives several worked-through experimental examples of the entanglement witnessing protocol.


\section{Definitions and equations \label{sec:Definitions}}

\textit{Quantum entanglement} is when two degrees of freedom cannot be described independently of each other: measuring one affects the other and vice versa.  Formally, this is defined as 
a state being non-separable \cite{bertlmann2023modern}. 
However, because such properties rely upon complex amplitudes and are thus not straightforwardly measurable, an alternative approach is to use indirect measures of entanglement, called \textit{entanglement witnesses}: observable quantities which detect the presence of entanglement in a system \cite{Bruss_2002,Guehne2009}. The most famous (and first)  entanglement witness is Bell's inequality, which reveals the entanglement between two particles separated in space \cite{PhysicsPhysiqueFizika.1.195,PhysRevLett.23.880}, versions of which have been experimentally verified to remarkable precision  \cite{PhysRevLett.28.938,Aspect_1999,Brunneretal2014}. 

However, Bell's approach requires individually manipulating and measuring particles. The problem becomes quite different when one considers the entanglement between electrons in solid materials, where $\sim 10^{23}$ electrons interact and cannot be individually manipulated and measured. 
In this case, it has been shown recently that spectroscopy can be used to witness solid state quantum entanglement \cite{Hauke2016,PhysRevResearch.2.043329,PhysRevB.103.224434,PhysRevLett.127.037201,Scheie2024_KYS,PontusEntanglementReview}. 
Here we use three experimentally validated entanglement witnesses, namely the Quantum Fisher Information, one-tangle, and two-tangle, to illustrate the method for extracting quantum information from neutron data---and importantly, how to use \textit{combinations} of entanglement witnesses to describe a system's quantum state. 

The definitions for these quantum entanglement witnesses are as follows: 


\subsection*{Quantum Fisher Information}

Quantum Fisher Information (QFI) gives a lower bound on  \textit{multipartite entanglement}, or the number of entangled objects which is also called \textit{entanglement depth} \cite{Hyllus2012, Toth2012}. 
For a pure state $| \psi \rangle$, the QFI of a system is directly proportional to the variance of an observable
\begin{equation}
    f_\mathcal{Q}  =  4 \left( \langle \psi |  \mathcal{O}^{\dagger}  \mathcal{O}  | \psi \rangle - \langle \psi | \mathcal{O} | \psi \rangle^2  \right) \label{eq:PureQFI}
\end{equation}
where $\mathcal{O}$ is a Hermitian quantum operator with a bounded eigenvalue spectrum \cite{Hauke2016}. 
For magnetic neutron scattering, $\mathcal{O} = S_{\alpha}({\bf Q}) =  \sum_{j} {S^{\alpha}}_j e^{i {\bf Q} \cdot {\bf r}_j}$.  
Readers familiar with neutron scattering will note that the first term in Eq. \ref{eq:PureQFI}, $\langle S_{\alpha}^{\dagger}({\bf Q}) S_{\alpha}({\bf Q})  \rangle$ corresponds to the total scattering at a particular wavevector $\bf Q$, and the second term $\langle S_{\alpha}({\bf Q})  \rangle^2$ corresponds to the elastic ($\hbar \omega = 0$) scattering at wavevector $\bf Q$ \cite{Squires}. Therefore, for a pure state ($T = 0$ limit), $f_\mathcal{Q}$ is proportional to the difference between total and elastic scattering. In other words, the energy-integrated inelastic scattering at $T=0$ corresponds to the QFI. 

Unfortunately, real experiments are always performed at nonzero temperature and typically 
probe thermally mixed states, not a pure state. A key breakthrough was provided by Hauke et al. \cite{Hauke2016} who proved that the equation for QFI of a thermal state is related to an energy integral over the imaginary part of the dynamic susceptibility:
\begin{equation}
	f_\mathcal{Q} \left[ {\bf Q}, T \right]  = \frac{4}{\pi}	\int_{0}^\infty \mathrm{d}(\hbar \omega) \tanh \left( \frac{\hbar \omega}{2 k_BT }\right) \chi^{\prime\prime}_{\alpha \alpha} \left( {\bf Q},  \omega\right),	\label{eq:QFI:Hauke}
\end{equation}
where for neutron scattering imaginary dynamic susceptibility is conventionally 
\begin{equation}
 \chi_{\alpha \beta}''\left( {\bf Q}, \omega \right)	=
 \pi \left( 1-e^{-\hbar\omega/k_B T} \right) S_{\alpha \beta}({\bf Q},\omega)	
 \label{eq:fluctdiss:DSFchi}
\end{equation}
\cite{Lovesey1984} 
and the neutron structure factor is 
\begin{equation}
    S_{\alpha \beta}({\bf Q},\omega) = \frac{1}{2 \pi \hbar} \int_{-\infty}^{\infty}  dt
    \langle \hat{S}^{\dagger}_{\alpha}({\bf Q})   \hat{S}_{\beta}({\bf Q}, t)  \rangle \exp (-i \omega t) 
\end{equation}
\cite{Squires}. 
Thus QFI can be experimentally determined at an arbitrary temperature. 
Note that as $T \rightarrow 0$, the tanh factor in Eq. \ref{eq:QFI:Hauke} approaches 1, Eq. \ref{eq:fluctdiss:DSFchi} becomes $\chi''\left( {\bf Q}, \omega \right) = \pi S \left( {\bf Q}, \omega \right) $, and thereby Eq. \ref{eq:QFI:Hauke} reduces to Eq. \ref{eq:PureQFI}.

The relationship between QFI and multipartite entanglement is 
$$f_\mathcal{Q} > m(h_{max} - h_{min})^2$$
where $m$ is an integer and is the lower bound on entanglement depth, and $h_{min/max}$ are the minimal and maximal eigenvalues of the local operators building up $\mathcal{O}$
\cite{Hyllus2012, Toth2012, PhysRevB.106.085110, Pezze2014}. 
The choice of $\mathcal{O} = S_{\alpha}({\bf Q})$ in Eq. \ref{eq:PureQFI} fixes  
 $h_{max} = - h_{min} = S$ such that $(h_{max} - h_{min})^2 = 4S^2$. Thus we can define normalized QFI (nQFI) as  
\begin{equation}
    \mathrm{nQFI}\left[ {\bf Q}, T \right]= \frac{f_\mathcal{Q}\left[ {\bf Q}, T \right]}{4 S^2} > m,
    \label{eq:nQFI}
\end{equation}
where $S$ is the quantum spin length and $m$ is an integer, and $\mathrm{nQFI} > m$ indicates a system with at least $m+1$ partite entanglement  \cite{Hauke2016,PhysRevB.103.224434} (e.g., $\mathrm{nQFI} = 2.718$ would indicate $\geq 3$-partite entanglement). In this way, a lower bound on entanglement can be probed as a function of temperature. 

For most practical purposes, Eqs. \ref{eq:QFI:Hauke}, \ref{eq:fluctdiss:DSFchi}, and \ref{eq:nQFI} can be combined into a single equation:
\begin{align}
\begin{split}
        {\rm nQFI}[{\bf Q},T] =  \frac{1}{S^2} & 
    \int_{0}^\infty   \mathrm{d}(\hbar \omega) \bigg[ \tanh \left( \frac{\hbar \omega}{2 k_BT }\right) \\ 
    &\left( 1-e^{-\hbar\omega/k_B T} \right) S_{\alpha \alpha}({\bf Q},\omega) 
    \bigg] \label{eq:bignQFI}
\end{split}
\end{align}
relating $S_{\alpha \alpha}({\bf Q},\omega)$ to nQFI, and thus to a lower bound on multipartite entanglement. 
Note that the bound in Eq. \ref{eq:nQFI} is derived for a single polarization component $S_{\alpha \alpha}$. For data where all components are summed, i.e. $S_{\rm tot} = S_{xx} +  S_{yy} +  S_{zz}$, the bound is instead $\mathrm{nQFI}\left[ {\bf Q}, T \right]= \frac{f_\mathcal{Q}\left[ {\bf Q}, T \right]}{12S^2} > m$ \cite{PhysRevLett.127.037201}.

\subsection*{\texorpdfstring{One-tangle $\tau_1$}{One-tangle tau1}}

The one-tangle \cite{PhysRevA.61.052306, PhysRevLett.96.220503} is a quantity measuring the entanglement between a $S=1/2$ degree of freedom and the rest of the system. As such, it allows powerful statements to be made about the quantum state of the system. It is defined at $T=0$  as 
\begin{align}
	\tau_1	& = 1-4\sum_\alpha \langle S_{j_0}^\alpha \rangle^2.
 \label{eq:one-tangle}
\end{align}
where $\alpha \in \{x,y,z \}$ and $j_0$ is the site index of an arbitrary site of the lattice. Note that $\langle S_{j_0}^\alpha\rangle$ represents a measurement on a single site or magnetic sublattice, i.e. it does not vanish in, for example, a classical N\'eel-ordered state.  
$\tau_1$ is essentially a measure of the static $T=0$ moment (as measured e.g. by the elastic magnetic scattering). Conceptually, it can be understood as follows: when a spin interacts with its environment, time reversal symmetry will be broken at the lowest temperatures, and the spin freezes. However, if there is nonzero quantum entanglement, spins will not completely freeze as $T \rightarrow 0$. The one-tangle is maximal ($\tau_1 = 1$) when a spin is completely dynamic and minimal ($\tau_1 = 0$) when it is completely static. 
In reality it is not possible to measure at $T=0$, so this measure should only be applied at low temperatures when it is clear that a system's order parameter is saturated. 

\subsection*{\texorpdfstring{Two-tangle $\tau_2$}{Two-tangle tau2}}

The two-tangle is a measure of total bipartite (pairwise) entanglement, which is based on concurrence \cite{PhysRevLett.80.2245}, which is itself a measure of entanglement between two objects. 
For a system with translational invariance, the concurrence between two $S=1/2$ spins is related to spin correlations via
\begin{align}
	C_{ij} &=	2\max \left\{ 0, \left| g^{xx}_{ij} - g^{yy}_{ij} \right| - \frac{1}{4} + g^{zz}_{ij},  \right.\nonumber\\
	&\left. \left| g^{xx}_{ij} + g^{yy}_{ij} \right| -\sqrt{\left( \frac{1}{4} + g^{zz}_{ij}\right)^2 -\left(M^z_{ij}\right)^2 }\right\}, \label{eq:concurrence}
\end{align}
where $g^{\alpha \alpha}_{ij} = \langle \hat{S}^\alpha_i \hat{S}^\alpha_j \rangle$ is the real-space two-point spin correlator between spins at sites $i$ and $j$, and 
$M^{\alpha}_{ij} = \frac{1}{2}(\langle \hat{S}^\alpha_i \rangle + \langle \hat{S}^\alpha_j \rangle)$
is the static magnetism~\cite{PhysRevA.69.022304}. 
The two-tangle is the sum of concurrences squared
\begin{equation}
    \tau_2  =   \sum_{i\neq j}  C^2_{ij}.
    \label{eq:two-tangle}
\end{equation}
This formulation is useful because the ratio $\tau_2/\tau_1$ gives the fraction of a system's entanglement that is pairwise~\cite{Amico_PRA_2006}. 

The requisite $g^{\alpha \alpha}_{ij} = \langle \hat{S}^\alpha_i \hat{S}^\alpha_j \rangle$ can be extracted from neutron scattering (which may require polarized experiments). Note however that a lattice average can severely affect the ability to measure $\tau_2$ for 2D and 3D lattices, where the sum over sites can mask large individual concurrence values in {\it e.g.} disordered systems \cite{ShannonAveraging2024}. Additionally, quantum monogamy when there are several neighbors for each site can result in the concurrence in each bond not being witnessed due to the threshold not being exceeded despite a fuly quantum state. Thus the two-tangle may be only experimentally useful for dimer or one-dimensional spin systems.

\section{Data analysis protocol \label{sec:Protocol}}

In this section we outline a step-by-step procedure for extracting quantum entanglement from neutron scattering data. 
Because it is the most involved, we begin with the procedure for calculating nQFI, which then serves as a template for other methods of computing entanglement.

\tikzstyle{input} = [rectangle, rounded corners, text width=1.2cm, minimum height=1.cm,text centered, draw=black, fill=green!25]
\tikzstyle{model} = [rectangle, rounded corners, text width=1.8cm, minimum height=1.cm,text centered, draw=black, fill=blue!16]
\tikzstyle{fit} = [rectangle, rounded corners, text width=1.7cm, minimum height=1.cm, text centered, draw=black, fill=orange!25]
\tikzstyle{output} = [rectangle, rounded corners, text width=1.0cm, minimum height=0.9cm,text centered, draw=black, fill=red!30]

\tikzstyle{arrow} = [thick,->,shorten >=0.1cm,shorten <=0.1cm]
\tikzstyle{curvedarrow} = [thin, dashed,->,shorten >=0.13cm,shorten <=0.13cm]

\begin{figure}
	\centering
	
	\begin{tikzpicture}[node distance=1.9cm]
	\node (crosssec) [input] {\large $\frac{d^2 \sigma}{d \Omega d\omega}$};
	\node (structfact0) [model, right of=crosssec, xshift=1.2cm] {$\tilde{S}({\bf Q},\omega)$};
	\node (structfact1) [model, right of=structfact0, xshift=1.4cm] {$S_{\alpha \alpha}({\bf Q},\omega)$};
 	\node (suscept) [model, below of=structfact1, yshift=-0.2cm] {$\chi''_{\alpha \alpha}({\bf Q},\omega)$};
	\node (QFI) [fit, below of=structfact0,   yshift=-0.2cm] {$f_{\mathcal{Q}}[{\bf Q}]$};
 	\node (nQFI) [fit, below of=crosssec,   yshift=-0.2cm] {${\rm nQFI} [{\bf Q}]$};
	
	\draw [arrow] (crosssec) -- (structfact0) node[midway,below] () {Eq.~\ref{eq:CrossSection}};
	\draw [arrow] (structfact0) -- (structfact1) node[midway,below]  () {Eq.~\ref{eq:PolarizationFactor}};
	\draw [arrow] (structfact1) -- (suscept) node[midway,right]  () {Eq.~\ref{eq:fluctdiss:DSFchi}};
	\draw [arrow] (suscept) -- (QFI) node[midway,below] () {Eq.~\ref{eq:QFI:Hauke}};
	\draw [arrow] (QFI) -- (nQFI) node[midway,below]  () {Eq.~\ref{eq:nQFI}};
 
	\draw [curvedarrow] (structfact1) .. controls (5,-1.2) and (1,-0.6)  .. (nQFI) node[midway,below]  () {Eq.~\ref{eq:bignQFI}};

	

 
	\end{tikzpicture}
	
	\caption{Workflow for calculating nQFI from neutron scattering data. The greatest experimental effort is converting $\frac{d^2 \sigma}{d \Omega d\omega}$ to $S_{\alpha \alpha}({\bf Q},\omega)$ (eqs. \ref{eq:CrossSection} and \ref{eq:PolarizationFactor}).}
	
	\label{flo:workflow}
\end{figure}
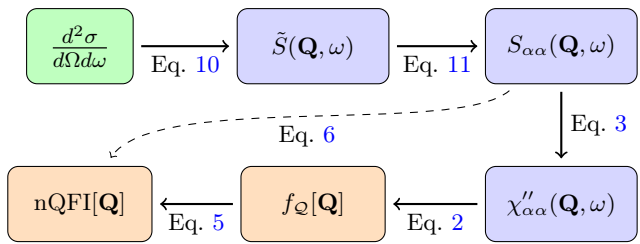

\subsection*{Quantum Fisher Information}

The workflow for converting a neutron scattering signal to a quantum entanglement bound is shown in Fig. \ref{flo:workflow}. 
Eq. \ref{eq:QFI:Hauke} for QFI is directly related to the neutron structure factor $S_{\alpha \alpha}({\bf Q},\omega)$.
However, a real neutron scattering experiment measures not $S_{\alpha \alpha}({\bf Q},\omega)$ directly, but the magnetic differential cross section
\begin{equation}
    \frac{d^2 \sigma}{d \Omega d\omega} = \frac{k_f}{k_i} N \left( \frac{ \gamma r_0}{2} g f({\bf Q}) \right)^2 e^{-2W} 
    \tilde{S}({\bf Q},\omega)
     \label{eq:CrossSection}
\end{equation}
\cite{Xu_AbsUnits_2013}
where $N$ is the number of magnetic atoms in the beam, $k_i$ and $k_f$ are the incident and scattered neutron wavevectors, $\gamma r_0 = 0.539 \times 10^{-14}$~m is a fixed value based on fundamental constants  \cite{Xu_AbsUnits_2013}, $g$ is the $g$-factor, $f({\bf Q})$ is the magnetic form factor, and $e^{-2W}$ is the Debye-Waller factor. 
$\frac{d^2 \sigma}{d \Omega d\omega}$ is in units of barn/(steradian$\cdot$meV). 
Meanwhile,  $\tilde{S}({\bf Q},\omega)$ is the spin structure factor modified by a polarization factor
\begin{equation}
    \tilde{S}({\bf Q},\omega) =  \sum_{\alpha, \beta} (\delta_{\alpha, \beta} - \hat{Q}_{\alpha} \hat{Q}_{\beta}) S_{\alpha \beta}({\bf Q},\omega)
     \label{eq:PolarizationFactor}
\end{equation}
where $\hat{Q}_{\alpha}$ and $\hat{Q}_{\beta}$ denote the projection of normalized scattering vector ${\bf Q}$ along the Cartesian axes $\alpha,\beta \in \{x,y,z \}$. 
Thus, to obtain $S_{\alpha \alpha}({\bf Q},\omega)$, one must convert the data to absolute units, exclude any non-magnetic scattering (e.g. phonon scattering or background artifacts), and correct for all the prefactors in Eq. \ref{eq:CrossSection}.  $S_{\alpha \alpha}({\bf Q},\omega)$ used for entanglement witnessing is thus in units of 1/energy (usually meV$^{-1}$) per magnetic ion. 

The explicit data analysis protocol for evaluating QFI from neutron scattering data is as follows:
\begin{enumerate}
    \item Isolate magnetic scattering by subtracting phonon scattering, sample holder scattering, etc.  
    
    \item Correct for the magnetic form factor and $g$-factor (and Debye-Waller factor if necessary, but this effect is weak at low $|Q|$ and low temperatures and can often be neglected). 
    
    \item If unpolarized neutron scattering is measured, correct for the polarization factor $(\delta_{\alpha \beta} - \hat{Q}_{\alpha} \hat{Q}_{\beta})$.  
    \\ Note that for isotropic magnetic systems (where the anisotropy is small enough to be neglected), this step is simplified as $S_{xx}({\bf Q},\omega) = S_{yy}({\bf Q},\omega) = S_{zz}({\bf Q},\omega) = \frac{1}{2} \tilde{S}({\bf Q},\omega)$, and the polarization factor is replaced by a factor of two. In anisotropic systems the individual polarization channels must be measured separately, see Section \ref{sec:ExperimentPlanning}. 
    
    \item Normalize data to absolute units, effectively determining $N$ in Eq. \ref{eq:CrossSection}. Many methods will provide such normalization, see Ref. \cite{Xu_AbsUnits_2013}. However, one must ensure the sum rule for spin length $S$  
    \begin{equation}
        \frac{\int_{-\infty}^{\infty} d\omega \int_{BZ} d{\bf Q} \sum_{\alpha}  S_{\alpha \alpha}({\bf Q},\omega) }{\int_{BZ} d{\bf Q}} = S(S+1) \label{eq:sumrule}
    \end{equation}
    is satisfied. This is especially important if one is dealing with significant orbital contribution to the ground state doublet and treating it as effective $J=1/2$ (e.g. Yb$^{3+}$ in KYbSe$_2$ \cite{Scheie2024_KYS}). Then to get an accurate QFI bound, the data should be normalized such that Eq. \ref{eq:sumrule} is consistent with  $S=1/2$ moments.
    
    \item Remove or mask the elastic line scattering. With finite energy resolution there is always a range of $\hbar \omega$ which is dominated by elastic scattering; it must be removed for an accurate QFI integral, especially when integrating at a wavevector of a magnetic Bragg peak. (For gapless systems this will suppress the QFI slightly, but it is a necessary cost to avoid overestimating the lower bound.) 
    
    \item Numerically evaluate the integral in Eq. \ref{eq:QFI:Hauke} up to the bandwidth of the magnetic scattering signal. 
\end{enumerate}

Two final things to keep in mind are (i) QFI can be evaluated for any wavevector $\bf Q$, but to get the highest bound on entanglement one should choose $\bf Q$ such that nQFI is maximal, typically where there is diverging intensity in the inelastic channel. 
(ii) QFI should be evaluated for a single component $\alpha \in \{x,y,z \}$ rather than summing them, as QFI as a bound in Eqs. \ref{eq:PureQFI} and \ref{eq:QFI:Hauke} to entanglement depth is only defined for a single operator $S_{\alpha}({\bf Q})$.

\subsection*{One and two-tangles}

The one-tangle in Eq. \ref{eq:one-tangle} is computed by first calculating the static magnetic moment $\langle M_{\alpha} \rangle$, and correcting for the $g$-factor as $\langle M_{\alpha} \rangle = g_{\alpha \alpha} \langle S_{\alpha} \rangle$. This is done either (i)  by doing a magnetic refinement, or (ii) by  summing the elastic magnetic scattering. 
The former is usually more precise but is only sensitive to long range order (not static magnetic disorder). The latter is sensitive to static disordered spins, but requires an absolute unit conversion via steps 1-4 of the QFI protocol above and is typically less precise (absolute unit conversions typically carry 20\% uncertainty  due to experimental limitations \cite{Xu_AbsUnits_2013}).  For systems with orbital contributions to magnetism, the one-tangle can be calculated as a fraction of the total effective moment (as determined by e.g. susceptibility or crystal field fits) \cite{Scheie2024_KYS}. 

The two-tangle in Eq. \ref{eq:two-tangle} is calculated from the spin correlations, which are extracted from neutron spectroscopy either through Fourier-transforming to real space \cite{PhysRevB.103.224434} or via sum rule analysis \cite{Xu_PRL_2000_CuN}. Either method requires following steps 1-4 of the QFI protocol above to obtain $S_{\alpha \alpha}({\bf Q},\omega)$ in absolute units. The Fourier transform route requires measuring a full Brillouin zone of scattering, whereas the sum rule analysis method does not---but the sum rule analysis assumes perfectly isotropic interactions, which is not always true. 
And thus $\tau_2$ is experimentally measurable using spectroscopy.

\section{Interpretation \label{sec:Interpretation}}

Each entanglement witness requires care in its interpretation. 
A first and important note is that the entanglement witnesses discussed here only probe entanglement between local spin degrees of freedom, not itinerant spins. (The QFI workflow can, however, be extended to electronic systems; see Refs. \cite{PontusEntanglementReview, PhysRevB.106.085110,fang2024amplified}.) 
Nor do they probe all forms of entanglement possible between local spins. Each witness can nevertheless reveal important information: 

\subsection{Quantum Fisher Information}

As a measure of entanglement depth, large QFI values can be used to rule out trivial un-entangled phases (as was done e.g. in triangular KYbSe$_2$ \cite{Scheie2024_KYS}). This is useful for distinguishing spectroscopy signals from classical glassy ground states and quantum disordered ground states, both of which have broad, diffuse features \cite{alexandradinata2020future}. 

However, the inverse is not true: a low nQFI does not rule out highly entangled behavior. Because the witness in Eq. \ref{eq:nQFI} is a \textit{lower bound}, one cannot infer anything about the underlying quantum state if  $\rm nQFI \leq 1$. For example, some topological phases are highly-entangled but require a nonlocal operator to witness rather than $\mathcal{O} = S_{\alpha}({\bf Q})$ \cite{PhysRevB.108.144414}. (In such cases one would expect intensity to be distributed over a large region of reciprocal space.) So just like one order parameter cannot witness all phase transitions, one entanglement witness cannot witness all types of entanglement \cite{PontusEntanglementReview}. 

Many have noted the connection between multipartite entanglement and quantum criticality \cite{Oliveira_2006_Multipartite,PhysRevA.82.062313,Liu_2013,PhysRevA.80.012318,PhysRevLett.119.250401,Hauke2016,Li_2024_multipartite,fang2024amplified,PhysRevD.96.126007}, and QFI especially is often used as a theoretical signature of quantum phase transitions \cite{PhysRevA.82.062313,Liu_2013,PhysRevA.80.012318,PhysRevLett.119.250401,Hauke2016,fang2024amplified}. 
Moreover, QFI has also been shown theoretically to have special scaling behavior with temperature around quantum critical points when $\mathcal{O}$ is related to a relevant order parameter \cite{Hauke2016,Gabbrielli2018}, and variations of QFI (e.g., quantum variance, where the $\tanh$ function in Eq. \ref{eq:QFI:Hauke} is replaced by a Langevin function  \cite{FrerotR2016}) have been used to explicitly define a quantum critical fan \cite{Frerot2019}.
Thus a key use of spectroscopic multipartite witnesses will be to evaluate the presence of, and a material's proximity to, a quantum phase transition. However, similar to the discussion above, not all quantum phase transitions are visible to QFI based on $S_{\alpha}({\bf Q})$, c.f. Cs$_2$CoCl$_4$ \cite{PhysRevLett.127.037201}.

\subsection{One and two-tangles}

The \textit{one-tangle} is simple to interpret: a reduced $T \rightarrow 0$ static moment indicates quantum entanglement.   
However, there are three important warnings in interpreting the one-tangle. The first is that the one-tangle does not reveal what kind of entanglement is present: it could be trivial singlet pair formation, or an exotic long ranged entangled phase. Both give large $\tau_1$ values. 
Second, $\tau_1$ is only defined for $S=1/2$. For larger spins, other mechanisms like multipolar order or single-ion singlets can reduce the order in the dipolar sector, but not because of entanglement with neighboring spins. 
Third, one must be careful not to conflate lack of long range ordered moment with a lack of static magnetism. Glassy or hidden order states \cite{paddison2015hidden} have static spins, even though no magnetic Bragg intensities are measurable. 
If these scenarios are not ruled out, one-tangle serves as an upper bound on the entanglement between one spin and the rest of the system. 

The \textit{two-tangle} is easier to interpret, as nonzero two-tangle along a given spatial direction indicates pairwise entanglement. As noted above, the ratio $\tau_2/\tau_1$ is an explicit measure of the fraction of entanglement that is pairwise \cite{Amico_PRA_2006}. 
However, for types of order where entanglement does not follow a specific spatial direction, the lattice average inherent in spectroscopy can suppress the measured $\tau_2$ to zero, even though microscopically nonzero concurrences may abound \cite{ShannonAveraging2024}. Thus the two-tangle may only be useful for one-dimensional systems. Even so, it can prove a useful one as shown by examples 1 and 2 below. 

\subsection{Combining multiple witnesses}

The greatest power in determining a material's quantum ground state is in combining multiple entanglement witnesses with conventional analysis (e.g., diffraction and refinement). Table \ref{tab:witnessInterpretation} shows interpretations of combinations of nQFI, $\tau_1$, and $\tau_2$. 
The combination of $\tau_1$ and nQFI is especially powerful: if both are large, this indicates a strongly fluctuating phase with entanglement distributed through the lattice. Meanwhile, if both are small, this points to a conventional magnetic state with a low degree of entanglement. 
Systematically probing many entanglement witnesses can offer valuable model-independent constraints on a given system's ground state. 
\begin{table}[]
    \centering
    \begin{tabular}{c|c|c|c}
        nQFI & $\tau_1$ & $\tau_2$  & Interpretation \\
        \hline \hline 
        large & large & - & extended entangled quantum state \\ \hline
        small & large & large & pairwise dimer entanglement \\ \hline
        small & large & - & local singlet formation or  entanglement  \\
          &  &   & inaccessible to two-point correlations \\ \hline
        large  & small & - & fluctuating magnetism close to instability \\ \hline
        small  & small & - & conventional low-entangled state \\
    \end{tabular}
    \caption{Interpretations of combinations of nQFI, $\tau_1$, and $\tau_2$ entanglement witnesses. $\tau_2$ is mainly useful for identifying dimer singlets which are periodic in the lattice. The dash - indicates when $\tau_2$ is excluded from the analysis.}
    \label{tab:witnessInterpretation}
\end{table}

As the field of spectroscopic entanglement witnesses continues to evolve, there will surely be additional constraints provided on the possible ground states. We hope that Table \ref{tab:witnessInterpretation} is merely the beginning of using entanglement witnesses to probe the unknown in quantum materials. 

\section{Planning the experiment \label{sec:ExperimentPlanning}}

The intermediate step to calculating entanglement from neutron data is calculating $S_{\alpha \alpha}({\bf Q},\omega)$. Therefore the experiment must be run with an eye toward background reduction (from the sample environment, sample mount, or the sample itself), eliminating phonons from the spectra (either through first principles or phenomenological modeling), and removing the tails of elastic line scattering from the inelastic channels. 
In addition, the experiment must have a means of normalizing the data to absolute units---such as vanadium standard measurements, normalization to phonons, or elastic incoherent scattering.

It is also important to have very good energy resolution around low-energy features at low temperature. This 
is because the hyperbolic tangent in Eq. \ref{eq:QFI:Hauke} is sensitive to finer and finer energy features as $T \rightarrow 0$. If these features are experimentally broadened in either momentum or energy, the measured QFI will be suppressed relative to the theoretical QFI without broadening. As a general rule, one is only sensitive to temperatures such that $k_B T > \Delta \hbar \omega$ (where $\Delta \hbar \omega$ is the energy resolution). 

Another important aspect of experimental planning is correcting for the polarization factor. As mentioned above, this is trivial for isotropic systems (at least above the ordering temperature), but for anisotropic systems a polarized spectrum should be measured. This can either be done with a spin-polarized time of flight spectrometer \cite{Zaliznyak_2017,Gabrys1997} or---if one knows exactly what wavevector to probe---with spin polarized triple axis spectrometry \cite{Shirane_Shapiro_Tranquada_2002}.  Alternatively, computational modeling can be used to correct for the polarization factor, see Ref. \cite{PhysRevLett.127.037201}. 
(If no means of correcting the polarization factor is available, the system must be treated as isotropic: the polarization factor will simply suppress the scattering along certain wavevectors, and the lower bound will be lower than the ideal QFI bound on entanglement, see e.g. Ref. \cite{PhysRevLett.127.037201} and example 3 below.)

The magnetic form factor is another experimental issue that must be handled carefully. Often an isotropic form factor is a reasonable approximation for low-$|Q|$ scattering~\cite{BrownFF}. But in some cases (e.g., $d$-electron compounds with orbital order \cite{Walters2009} or hybridization effects \cite{Sarkis_2024}) an anisotropic form factor is necessary to account for covalent or orbital effects  \cite{boothroyd2020principles}.

It is also important to note that all of these entanglement witnesses assume well-defined spin degrees of freedom. If excited orbital or crystal field states are observable in the spectroscopy, they must be excluded from the analysis. If the total quantum number $S$ is not constant across the relevant bandwidth, it is not valid to apply an entanglement witness using the low-energy $S$ value. 

Finally, it is often advantageous to collect a full magnetic  Brillouin zone for this analysis. This is for three reasons: first, it allows one to cross-check the normalization to see that the sum rule Eq. \ref{eq:sumrule} is satisfied. Second, it allows one to calculate QFI for any wavevector ${\bf Q}$, to ensure that the maximal lower bound on entanglement depth is reported. Third, when the full Brillouin zone is measured, other quantum information quantities become available (e.g., real-space non-commutativity \cite{Scheie2022} or quantum coherence \cite{scheie2023reconstructing}).

\section{Examples \label{sec:Examples}}

{

\subsection*{Trivial un-entangled examples:}

As a preliminary, we first calculate the quantum entanglement for two trivial theoretical examples: a non-interacting $S=1/2$ paramagnet, and a one-dimensional ferromagnet. Both of these show no entanglement. 

\paragraph{Non-interacting $S=1/2$ paramagnet:} 
A non-interacting $S=1/2$ paramagnet is an array of spins which are all in the superposition state $| \psi \rangle = \frac{1}{\sqrt{2}}(|\uparrow\rangle \pm  |\downarrow\rangle )$, where the quantization axis is randomized on each site. In the local coordinate system, each spin thus has $\langle S^x_{j_0}\rangle =\pm 1/2$, $\langle S^y_{j_0}\rangle = \langle S^z_{j_0}\rangle =0$, resulting in a vanishing one-tangle $\tau_1=0$. 
The two-tangle $\tau_2 = 0$  as there is no correlation between any spins and all $g_{ij} = 0$. 
The pure state QFI meanwhile can be directly calculated from Eq. \ref{eq:PureQFI}, which gives $\rm nQFI=1$ for all polarization channels.  
(Although without interactions this system has no energy scale, this is consistent with the spectroscopic QFI: 
The neutron cross section of an ideal paramagnet is 
$
    S_{\alpha \alpha}({\bf Q}) =  \frac{1}{3}S(S+1) = \frac{1}{4}  
$ \cite{Squires}, which via Eq. \ref{eq:bignQFI} gives 
$
{\rm nQFI}[{\bf Q},T=0] = 1 
$ assuming all spectral weight is inelastic---an admittedly ill-defined concept for a system with no energy scale.) Thus no entanglement is witnessed by any measure, as the individual spins do not interact with each other. Note that for $T>0$ ${\rm nQFI}\rightarrow0$ as $k_BT$ will exceed the (quasi)elastic energy band of the paramagnet.  

\paragraph{One-dimensional ferromagnet:}
The one-dimensional Heisenberg ferromagnetic spin chain is a textbook spin wave problem, and the linear spin wave theory $T=0$ neutron cross section for spins polarized along $z$ is given by 
\begin{equation}
    S_{xx}({\bf Q}, \omega) = S_{yy}({\bf Q}, \omega) =  \frac{S}{2} \delta(\hbar \omega - \hbar \omega_{\bf Q})  \label{eq:1DHeisenberg}
\end{equation} 
where $\hbar \omega_{\bf Q}$ is the spin wave dispersion \cite{boothroyd2020principles}. 
The one-tangle $\tau_1$ is calculated from the static ordered moment $\langle S^z \rangle = \frac{1}{2}$ for the ideal ferromagnet \cite{boothroyd2020principles}, which via Eq. \ref{eq:one-tangle} gives $\tau_1 = 0$. 
The two-tangle is calculated from the Fourier-transform of energy-integrated Eq. \ref{eq:1DHeisenberg}, and yields $g^{zz}_{ij} = \frac{1}{4}$ and $g^{xx}_{ij} = g^{yy}_{ij} = 0$, which via Eq. \ref{eq:concurrence} gives $C_{ij} = 0$ for all $i$ and $j$, and thus  $\tau_2 = 0$. 
Meanwhile, the nQFI for $S=1/2$ and $T=0$ via Eqs. \ref{eq:1DHeisenberg} and \ref{eq:bignQFI} is 
$
{\rm nQFI}[{\bf Q},T=0] =  1 
$, 
which witnesses no multipartite entanglement. 
Although this model system is one-dimensional and would have no static magnetic order at $T>0$, the same conclusions generically hold for ordered ferromagnets in higher dimensions where Eq. \ref{eq:1DHeisenberg} describes the magnon intensity. 

For both the ideal paramagnet and the 1D Heisenberg ferromagnet, neither one-tangle nor two-tangle give any entanglement, and zero temperature nQFI is precisely one (a consequence of inelastic intensity equally distributed through the Brillouin zone). 
Although these two theoretical examples have very different 
Hamiltonians, they 
host product ground states with no witnessed entanglement. 

We now turn to more complex examples with experimental data:  

}

\subsection*{Experimental Example 1: Copper Nitrate}

\begin{figure}
    \includegraphics[width=0.98\columnwidth]{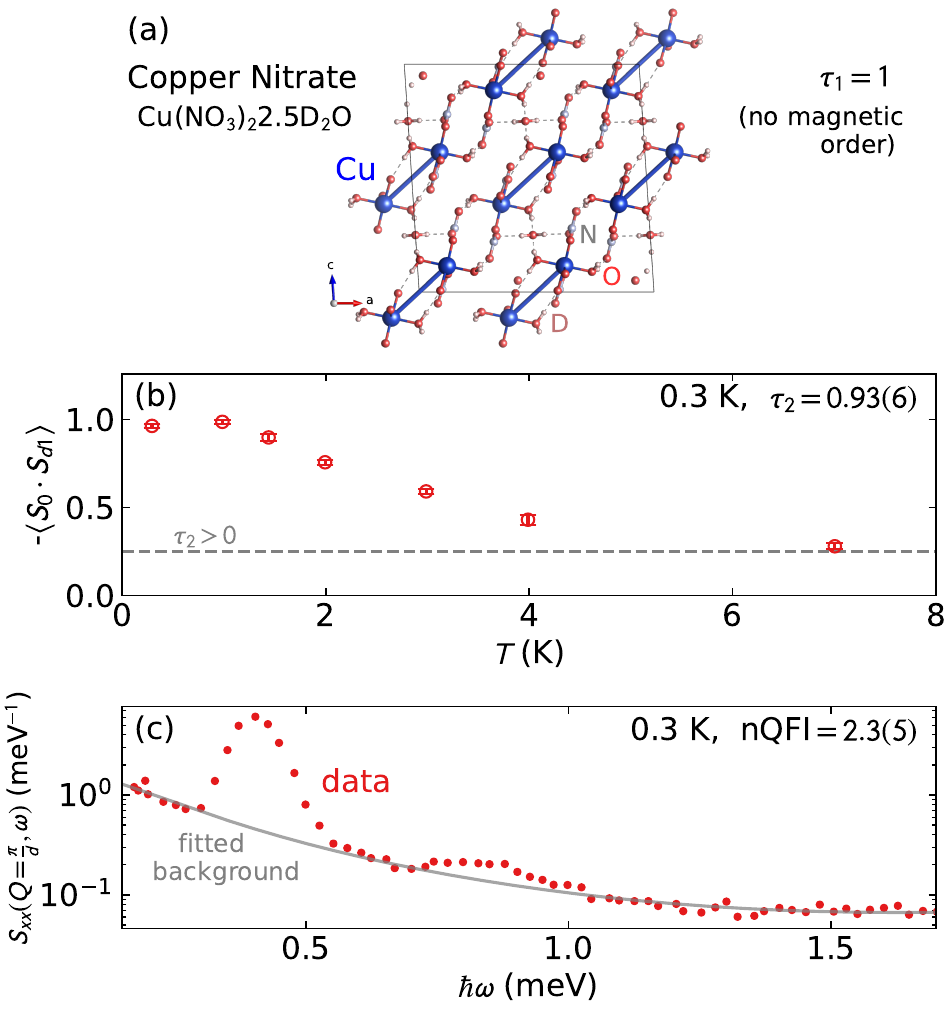} 
    \caption{Entanglement on Copper Nitrate measured by neutron scattering. Panel (a) shows the crystal structure \cite{Morosin:a07535}, where copper atoms form dimers. Panel (b) shows the nearest neighbor pairwise correlation measured in \cite{Xu_PRL_2000_CuN}. Panel (c) shows the 0.3~K inelastic spectra measured at the dimer distance from Ref. \cite{Tennant_PRB_2003_CuN}, with calculated nQFI=2.3(5).}
    \label{fig:CuN}
\end{figure}

Copper nitrate, Cu(NO$_3$)$_2$2.5D$_2$O, is a well-studied dimer system \cite{Morosin:a07535} where $S=1/2$ copper ions form quantum singlets at the lowest temperatures (Fig. \ref{fig:CuN}) and have well-defined excitations measured by neutron scattering \cite{Xu_PRL_2000_CuN,Tennant_PRB_2003_CuN}. 
It serves as an instructive example for how entanglement witnesses give a unique and clear perspective of its quantum ground state. 

\subsubsection{Theoretical values:} 
For comparison, let us first calculate the entanglement values for the idealized theoretical dimer. The theoretical structure factor for an isolated Heisenberg $S=1/2$ dimer at $T=0$ is 
\begin{equation}
    S_{\alpha \alpha}({\bf Q},\omega) =  \frac{1}{2} \sin^2({\bf Q}\cdot{\bf d}/2) \> \delta(\hbar \omega - J) 
    \label{eq:TheoreticalDimer}
\end{equation}
\cite{Zheludev_1996} where $\bf d$ is the intra-dimer separation vector,  $\delta(x)$ is the Dirac delta function, and $J$ is the intra-dimer exchange energy. 
The \textit{one-tangle} is trivial because there is no static magnetism, and so $\tau_1=1$. 

The \textit{two-tangle} can be derived from the Fourier transform of Eq. \ref{eq:TheoreticalDimer}, (or a simple first principles calculation) which gives $ g^{xx}_{1,2} = g^{yy}_{1,2} = g^{zz}_{1,2} = -1/4$, which via eq. \ref{eq:concurrence} gives $C_{1,2}=1$ and thus  $\tau_2 = 1$ via eq. \ref{eq:two-tangle}. 

The \textit{Quantum Fisher Information} is calculated from the maximum intensity in Eq. \ref{eq:TheoreticalDimer}, $S_{\alpha \alpha}({\bf Q},\omega) =  \frac{1}{2}$ at the intradimer separation wavevector $Q = \frac{\pi}{d}$, which via Eq. \ref{eq:bignQFI} yields $\rm nQFI = 2$ at $T=0$, which witnesses bipartite entanglement. 
If one accounts for the coupling between dimers in copper nitrate via perturbation theory, this theoretically increases the intensity at $Q = \frac{\pi}{d}$ by approximately 1.163 \cite{Tennant_PRB_2003_CuN} for a theoretical $\rm nQFI \approx 2.326$. 
This slight increase witnesses three-partite entanglement, indicating more extended entanglement through the lattice when dimers are coupled. 

\subsubsection{Experimental values:}  
We begin with the simplest witness: \textit{one-tangle}. Because there is no magnetic order at the lowest temperatures in copper nitrate \cite{Friedberg_1968}, the one-tangle is simply $\tau_1=1$. 

The \textit{two-tangle} can be calculated from concurrence (which has been done in Ref. \cite{PhysRevA.73.012110} from the spin correlations extracted from sum rule analysis \cite{Xu_PRL_2000_CuN}). Because of experimental artifacts and uncertainties, the measured dimer spin correlations plotted in Fig. \ref{fig:CuN}(b) actually exceed the physical bound $|\langle {\bf S}_i \cdot {\bf S}_j \rangle| \leq 3/4$ at low temperatures, which via equations \ref{eq:concurrence} and \ref{eq:two-tangle} give an unphysical $\tau_2 > 1$.  
We estimate the real $\tau_2$ by normalizing the largest measured $|\langle {\bf S}_i \cdot {\bf S}_j \rangle|$ to be $3/4$, which gives $\tau_2 = 0.93(6)$ at 0.3~K. This value should have larger uncertainty in reality because of the \textit{ad-hoc} normalization, but it is clearly something close to 1. 

Finally, the \textit{Quantum Fisher Information} is calculated from the energy dependent scattering at $Q = \frac{\pi}{d}$ (dimer coupling wavevector) reported in Ref. \cite{Tennant_PRB_2003_CuN}, plotted in Fig. \ref{fig:CuN}(c). Using the absolute unit conversion in Ref. \cite{Xu_PRL_2000_CuN}, we subtract the background as in Ref. \cite{Tennant_PRB_2003_CuN} and calculate an $\rm nQFI = 2.3(5)$ at 0.3~K via Eq. \ref{eq:bignQFI}. This witnesses $\geq 2$-partite entanglement, and is consistent with $\geq 3$-partite entanglement. This value is within error bars of the 2.326 theoretical nQFI value. 
The python code for the above data analysis is found at \url{github.com/asche1/EntanglementWitnessesTutorial}.

\paragraph{Interpretation:}

The experimental ratio $\tau_2/\tau_1 = 0.93(6)$ means that almost all the entanglement is pairwise \cite{Amico_PRA_2006}, and is very close to the theoretical $\tau_2/\tau_1 = 1$, consistent with dimer formation (see Section \ref{sec:Definitions}). 
Also, the  $\rm nQFI \geq 2$ means that QFI witnesses more than just bipartite entanglement, consistent with coupling between the dimers for a collective quantum state. 
Thus, if nothing were known about Cu(NO$_3$)$_2$2.5D$_2$O other than the entanglement witnesses reported above, we could confidently conclude that the ground state involves coupled pairwise dimer 
(or valence bond) 
formation. 

\subsection*{\texorpdfstring{Experimental Example 2: KCuF$_3$}{Example 2: KCuF3}}

 KCuF$_3$ is a quasi-one-dimensional antiferromagnetic spin chain system where, due to orbital order, $S=1/2$ copper atoms interact strongly along the $c$ axis and weakly in the $ab$ plane \cite{PhysRevLett.111.137205}. The entanglement witnesses of KCuF$_3$ were analyzed in detail in Ref. \cite{PhysRevB.103.224434}; here we summarize the results. 

We calculate the \textit{one-tangle} from the ordered moment refined from diffraction.  KCuF$_3$ has a 4~K ordered moment  of 0.49(7)~$\mu_B$ where the order parameter is saturated from an ordering temperature $T_N=39$~K \cite{Hutchings_1969}. Assuming $g=2.0$, this means $\langle S^z \rangle = 0.24(3)$ and the one-tangle is $\tau_1 = 0.76 \pm 0.14$. This is close to 1, indicating strong quantum entanglement. We note that this is a somewhat conservative estimate, and that slightly higher $\tau_1$ values can be obtained using experimental $g$ factors \cite{Miike1975}.

\begin{figure}
    \includegraphics[width=0.96\columnwidth]{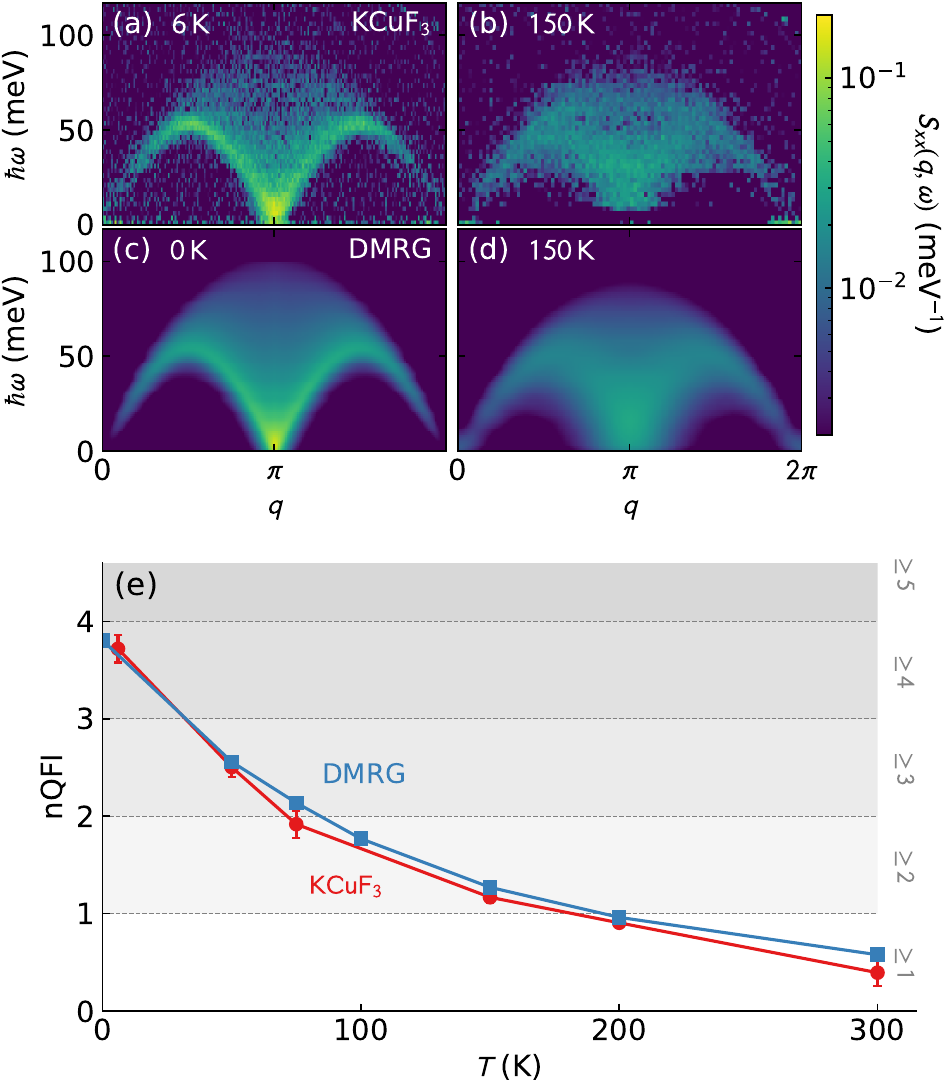} 
    \caption{Neutron spectra of spin chain KCuF$_3$ (a)-(b), DMRG simulation of the 1D Heisenberg chain (c)-(d), and calculated nQFI (e). Data from Ref. \cite{PhysRevB.103.224434}. Each integer $m$ nQFI indicates $\geq m+1$ partite entanglement.}
    \label{fig:KCuF3}
\end{figure}

The \textit{two-tangle} is derived by Fourier-transforming the energy-integrated neutron spectra (from in Fig. \ref{fig:KCuF3}) to get the spin correlations as a function of distance along the chain. At the lowest measured temperature 6~K, $\tau_2 = 0.16 \pm 0.03$. Comparison to Density Matrix Renormalization Group (DMRG) simulations showed this value is heavy influenced by experimental artifacts \cite{PhysRevB.103.224434}; 
the $T=0$ theoretical simulations produce the value 0.256. 

The \textit{Quantum Fisher Information} we calculate as a function of temperature by using Eq. \ref{eq:bignQFI} with $S=1/2$, evaluated at $Q=\pi$ along the chain (where the intensity is strongest). The results are shown in fig. \ref{fig:KCuF3}. At $T=6$~K, nQFI$=3.72 \pm 0.14$, witnessing at least four-partite entanglement per spin. Comparison to Bethe Ansatz calculations \cite{Caux_2009} on a 500-site chain shows this value would be around 20\% higher without experimental resolution broadening \cite{PhysRevB.103.224434}.

\paragraph{Interpretation:}

All three entanglement witnesses give some appreciable entanglement. The ratio $\tau_2/\tau_1 = 0.21(5)$ (or using the DMRG $\tau_2$ result, $\tau_2/\tau_1 = 0.34(6)$) shows that a minority of entanglement is pairwise. Meanwhile, the large nQFI shows appreciable entanglement along the chain. All this is consistent with the expected behavior of a 1D Heisenberg antiferromagnet, which is known theoretically to have a highly entangled ground state \cite{Laflorencie2016}. 
Furthermore, because of the noted connection between QFI and quantum criticality, the large nQFI (i.e., witnessing more than bipartite entanglement) indicates that  KCuF$_3$  is indeed quantum critical, as scaling analysis also shows \cite{Lake2005}. 

\subsection*{\texorpdfstring{Experimental Example 3: NiPS$_3$}{Example 3: NiPS3}}

We turn to our third experimental example NiPS$_3$ to illustrate (i) how resolution and domain effects can impact witnessed entanglement, and (ii) how magnons alone can produce nontrivial QFI. 

NiPS$_3$ is a $S=1$ quasi-2D honeycomb van der Waals antiferromagnet, which orders in a stripe pattern \cite{Wildes_2015_NiPS3}. It has a strongly reduced magnetic moment, and its excitations include anomalous intensity at low energies which linear spin wave theory (LSWT) models fail to account for \cite{Wildes_2022_NiPS3,Scheie_2023_NiPS3}. 
Because NiPS$_3$ is a $S=1$ system, the one and two-tangles (at least as formulated above) are not applicable and we focus on QFI.

\begin{figure*}
    \includegraphics[width=0.96\textwidth]{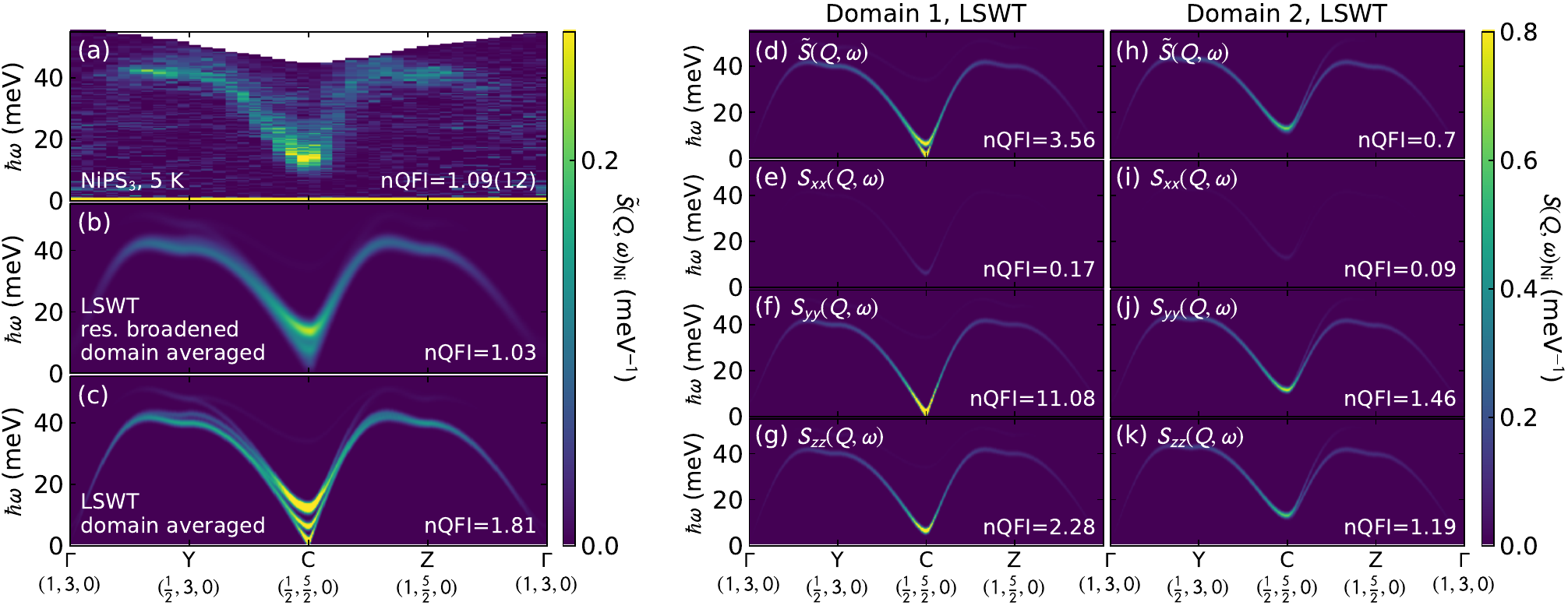} 
    \caption{NiPS$_3$ neutron spectra and Quantum Fisher Information evaluated at ${\bf Q} = (1/2, 5/2, 0)$ for various LSWT models. Panel (a) shows the experimental spectra and nQFI. Panels (b) and (c) show the domain-averaged LSWT with and without $Q$-resolution broadening (see Ref. \cite{Scheie_2023_NiPS3}). Panels (d)-(g) and (h)-(k) show the spectra from domains 1 and 2 respectively, both with the polarization factor (top) and the individual polarization channels $S_{xx}(Q,\omega)$, $S_{yy}(Q,\omega)$, $S_{zz}(Q,\omega)$. Note that measuring domain 1 unpolarized would yield $\rm nQFI > 1$ with $\geq 2$-partite entanglement; but resolving  $S_{yy}(Q,\omega)$ yields the largest  $S_{yy}(Q,\omega) > 3$, indicating $\geq 4$-partite entanglement. However, experimental effects suppress the nQFI just to the boundary of being able to witness nontrivial entanglement. All LSWT QFI calculations assume $T=0$.}
    \label{fig:NiPS3}
\end{figure*}

Using the data reported in Ref. \cite{Scheie_2023_NiPS3}, we calculate the nQFI using Eqs. \ref{eq:QFI:Hauke} and \ref{eq:nQFI}, choosing ${\bf Q} = (1/2, 5/2, 0)$ as the wavevector of greatest experimental intensity, shown in Fig. \ref{fig:NiPS3}. (Note that the data must be first normalized to intensity per Ni ion, whereas data in Ref. \cite{Scheie_2023_NiPS3} were reported as intensity per unit cell.) Because the measured data are unpolarized, we assume isotropic spins and $S_{\alpha \alpha}({\bf Q},\omega) = \frac{1}{2} \tilde{S}({\bf Q},\omega)$. 
Thus the equation relating nQFI to $S({\bf Q}, \omega)$ for NiPS$_3$ is 
\begin{align}
\begin{split}
        {\rm nQFI}({\bf Q}) = \frac{1}{2}
    \int_{0}^\infty  & \mathrm{d}(\hbar \omega) \bigg[ \tanh \left( \frac{\hbar \omega}{2 k_BT }\right) \\ 
    &\left( 1-e^{-\hbar\omega/k_B T} \right) \tilde{S} ( {\bf Q}, \hbar \omega ) 
    \bigg],
\end{split}
\end{align}
where the prefactor is from Eqs. \ref{eq:QFI:Hauke},  \ref{eq:fluctdiss:DSFchi}, and \ref{eq:nQFI} with $S=1$, and the isotropic polarization approximation. 
The experimental NiPS$_3$ data [Fig. \ref{fig:NiPS3}(a)] give $\rm nQFI = 1.09 \pm 0.12$ which agrees with the LSWT model calculation   $\rm nQFI = 1.03$ [Fig. \ref{fig:NiPS3}(b)]---just above the threshold for witnessing nontrivial entanglement, but within uncertainty of witnessing no entanglement. 
However, examining the LSWT simulations with \textit{SpinW}~\cite{SpinW} show that this value would be ten times larger without resolution and experimental effects. 

NiPS$_3$ is actually a slightly distorted honeycomb lattice with three domains which are averaged in a real experiment \cite{Scheie_2023_NiPS3}. Furthermore, the spectrometer ${\bf Q}$-resolution effects are significant at the bottom of the dispersion where the intensity is strongest \cite{Scheie_2023_NiPS3}. 
Although these effects are unavoidable experimentally, they can be un-done in simulations. In Fig. \ref{fig:NiPS3}(c) we turn off the finite momentum resolution, and find the nQFI nearly doubles to 1.81. In  Fig. \ref{fig:NiPS3}(d) and (h) we separate two domains, and find that in domain 1 the nQFI doubles again to 3.56---crossing the threshold into witnessing four-partite entanglement. 

This QFI analysis assumed isotropic polarization scattering. In reality, because of spin-ordering and planar anisotropy, the polarization channels are not identical. In Fig. \ref{fig:NiPS3}(e)-(g) we plot the different polarization channels $S_{xx}$,  $S_{yy}$, and  $S_{zz}$. In this case, we find that the  $S_{yy}$ channel in domain 1 (where the softest and most intense oscillation mode appears) gives $\rm nQFI = 11.08$, which is three times the nQFI from the raw unpolarized spectrum, witnessing at least 12-partite entanglement in NiPS$_3$. 

\begin{figure}
    \includegraphics[width=0.96\columnwidth]{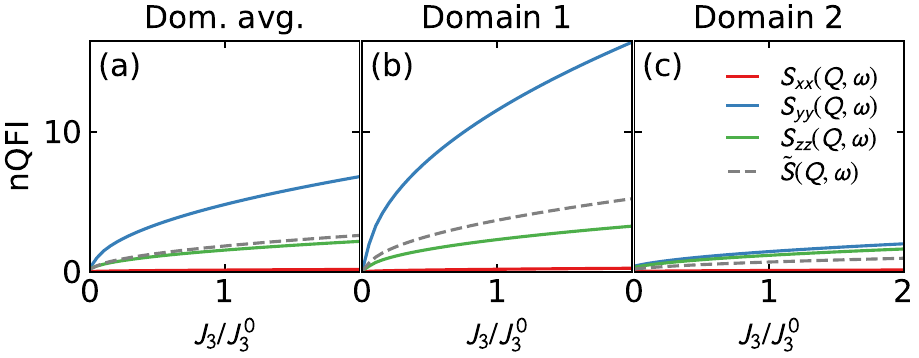} 
    \caption{NiPS$_3$ LSWT entanglement at ${\bf Q} = (1/2, 5/2, 0)$ and $T=0$ as the Hamiltonian is modified. As $J_3$ increases relative to the experimental fitted $J_3^0$, the nQFI entanglement increases. This is because the gap relative to the bandwidth decreases, increasing the intensity at ${\bf Q} = (1/2, 5/2, 0)$, and enhancing the nQFI.}
    \label{fig:NiPS3_trends}
\end{figure}

We can go a step further and then modify the Hamiltonian by increasing the third neighbor exchange $J_3$ (the largest exchange term in the Hamiltonian) and calculate the entanglement from LSWT, shown in Fig. \ref{fig:NiPS3_trends}. 
Across all sectors, but especially for $S_{yy}$ in domain 1, enlarging $J_3$ increases the QFI. This is because increasing $J_3$ increases the magnon bandwidth, decreasing the gap size relative to the bandwidth, increasing the intensity at the bottom of the dispersion, and increasing the witnessed QFI. 
On a very coarse level, this is consistent with QFI as a witness of quantum critical behavior: it is especially sensitive to a buildup of low-energy density of states at a particular wavevector. 

A caveat to all this is that the LSWT model used here fails to completely match NiPS$_3$ experiments \cite{Scheie_2023_NiPS3}. Nevertheless, the analysis above shows that for a spectrum like NiPS$_3$, resolution effects, domain averaging, and the polarization factor each individually suppress the nQFI by a factor of two or more---such that together they suppress the measured nQFI by a full order of magnitude ($\rm nQFI = 1.09$ experimentally vs $\rm nQFI = 11.08$ ideally). 
It also shows that nontrivial QFI values do not necessarily indicate non-magnon behavior. LSWT, although often thought of as a semiclassical method as it leaves out important corrections, is a quantum theory of noninteracting bosons (magnons). It can thus capture entanglement between spins---but not effects on the entanglement driven by magnon-magnon interactions, such as quantum renormalization, decay, etc. (Put another way, the very large entanglement depth is a consequence of antiferromagnetism where the Boglioubov transformation mixes the plane-wave states that spread over a wide number of sites on the different magnetic sublattices.) 
The example of NiPS$_3$ also shows that as magnon dispersions approach zero energy, the entanglement can grow to large values, consistent with antiferromagnetic fluctuations increasing as mode gaps go to zero. 
Thus even conventional magnon systems can yield large and nontrivial multipartite entanglement.  

\section{Conclusion}

We have described in detail the definitions and methodology for three entanglement witnesses applicable to magnetic neutron scattering. 
We focus on these three not because they are the only or most important witnesses, but  because  (a) they have been experimentally validated and their pitfalls are understood, (b) they serve as a helpful introduction to the field, and (c) the concepts we describe (especially the experimental considerations) apply to other entanglement measures as well. 
As far as the experimentalist is concerned, several features are in common between all spectroscopic entanglement witnesses: (1) the importance of an absolute unit scale for scattering intensity, (2) the importance of correcting for and eliminating experimental artifacts, and (3) the importance of using multiple entanglement witnesses to give a more complete picture of a system's underlying quantum state. 

We anticipate that many more experimental entanglement witnesses will emerge as the field progresses. 
However the methodology we outline here serves as an introduction to the field of spectroscopic entanglement witnesses and a template for other witnesses. 
As these examples show, there is a wealth of quantum information embedded in the spin correlation functions measured by spectroscopy; and with the right tools, such information can be experimentally accessed.

\section*{acknowledgments}

 The work by A.S., W.S., and D.A.T. is supported by the Quantum Science Center (QSC), a National Quantum Information Science Research Center of the U.S. Department of Energy (DOE). 
 The work of P.L. and E.D. was supported by the U.S. Department of Energy, Office of Science, Basic Energy Sciences, Materials Sciences and Engineering Division. 
 Final parts of the work by A.S. were performed at Aspen Center for Physics, which is supported by National Science Foundation grant PHY-2210452. 
 We are grateful to Je-Geun Park for many discussions on NiPS$_3$.

\end{document}